\documentclass[aps,prl,twocolumn,showpacs,a4paper,unsortedaddress,
amsmath,amssymb,byrevtex]{revtex4}

\usepackage[latin1]{inputenc}
\usepackage{graphicx}
\usepackage{dcolumn}
\usepackage{bm}
\usepackage{hyperref}
\usepackage{times}

\begin{document}
\preprint{ffuov/02-01}

\title{Conformation dependence of molecular conductance:
chemistry versus geometry}

\author{C. M. Finch}
\author{S. Sirichantaropass}
\author{S. W. Bailey}
\author{I. M. Grace}
\author{V. M. Garc\'{\i}a-Su\'arez}
\email{v.garcia-suarez@lancaster.ac.uk}
\author{C. J. Lambert}
\affiliation{Department of Physics, Lancaster University,
Lancaster, LA1 4YB, U. K.}

\date{\today}

\begin{abstract}
Recent experiments by Venkatamaran {\em et al.} [Nature (London)
{\bf 442}, 904 (2006)] on a series of molecular wires with varying
chemical compositions, revealed a linear dependence of the
conductance on $\mathrm{cos}^2\theta$, where $\theta$ is the angle
of twist between neighboring aromatic rings. To investigate
whether or not this dependence has a more general applicability,
we present a first principles theoretical study of the transport
properties of this family of molecules as a function of the
chemical composition, conformation and the contact atom and
geometry. If the Fermi energy $E_\mathrm{F}$ lies within the
HOMO-LUMO gap, then we reproduce the above experimental results.
More generally, however, if $E_\mathrm{F}$ is located within
either the LUMO or HOMO states, the presence of resonances
destroys the linear dependence of the conductance on
$\mathrm{cos}^2\theta$ and gives rise to non-monotonic behaviour
associated with the level structure of the different molecules.
Our results suggest that the above experiments provide a novel
method for extracting spectroscopic information about molecules
contacted to electrodes.
\end{abstract}


\maketitle

Single-molecule electronics poses many fundamental challenges for
chemistry, physics and engineering, partly because a molecule
attached to metallic electrodes is a quantum object that lives
between the traditional disciplines of chemistry and physics.
Challenges also arise, because many electrical properties of
single-molecule devices are exponentially sensitive to changes in
the environment and to details of the contact to electrodes. As a
consequence, experiments using mechanically-controlled break
junction techniques (MCBJ) \cite{Ree97} and scanning tunneling
microscopy (STM) \cite{Joa95,Dat97} yield broad distributions of
measured transport properties, corresponding to slightly different
attachments of molecules and in some cases, to environmental
fluctuations \cite{Che06,Lon06}. Given these sensitivities, it is
remarkable that systematic experimental studies can yield clear
trends associated with generic molecular features. One such
example is a study of the effect of tilting the angle of contact
between the long axis of a molecular wire and the plane of the
locally-flat electrode \cite{Hai06}. Another example is a recent
study of single-molecule electron transport as a function of
molecular conformation \cite{Ven06}.

\begin{figure}
\includegraphics[width=\columnwidth]{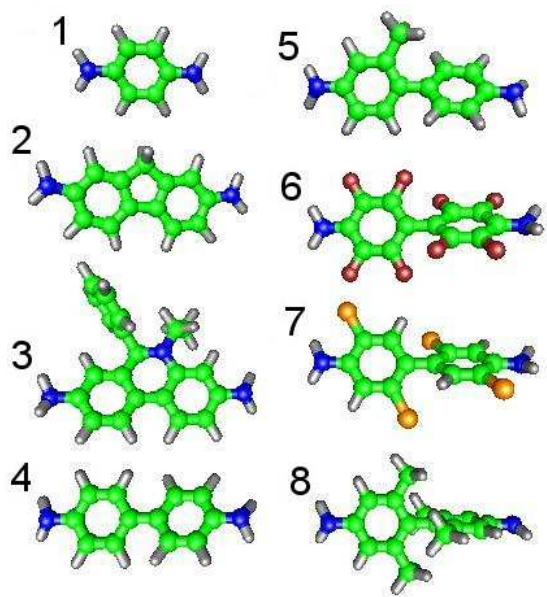}
\caption{\label{Fig1} GGA-relaxed configurations of all molecules
studied in this work capped with NH$_2$. Grey, green, blue, red
and orange vertexes correspond to H, C, N, F and Cl atoms,
respectively.}
\end{figure}

While both of the above studies reveal that geometry plays an
important role in controlling electron transport through single
molecules, the experiments of Ref. \onlinecite{Ven06} contain an
even stronger suggestion, namely that for the set of molecules
studied, changes in chemical composition are irrelevant. In these
experiments, each molecule in the chosen series, (see Fig.
(\ref{Fig1})), possessed two phenyl rings linked by a single
carbon-carbon bond. Different chemical groups on the molecules
caused the static angle of twist $\theta$ to vary from molecule to
molecule. The highest conjugation and therefore the highest
conductance is expected when the two rings lie in the same plane
(ie $\theta = 0$). However, when the rings are rotated relative to
each other, the  overlap between the $\pi$ orbitals decreases and
the electrical conductance follows a $\mathrm{cos}^2\theta$ law.
This simple dependence on geometry is remarkable, since different
chemical side groups were used to produce different angles. In
other words, it appears that chemistry is irrelevant and geometry
is everything.

In this article, we aim to understand whether or not the dominance
of geometry over chemistry can be widely expected or if this is a
peculiar property of the set of molecules measured. Close
inspection of the measured conductances in Ref. \cite{Ven06}
reveals that they do not follow a perfect $\mathrm{cos}^2\theta$
law and it is of interest to understand these deviations. The
influence of the chemical composition can be separated in two
parts. On the one hand, different atoms or molecular groups can be
attached to the aromatic rings to produce a desired angle. These
include hydrogen atoms, alkanes, other aromatic chains, nitrogen,
fluorine and chlorine atoms. Such groups define the ground-state
angle $\theta$, which can vary between 0 and almost 90 degrees. On
the other hand, it is also possible to employ different end atoms,
which attach the molecules to the leads. In the experiments,
Venkatamaran {\em et al.} \cite{Ven06,Ven06b} used molecules
capped with either thiol (SH) or amino groups (NH$_2$) and found
that nitrogen termination gave more reproducible results.

In what follows, we present a theoretical analysis of the
structural configurations and transport properties of a series of
molecules shown in Fig. (\ref{Fig1}), with thiol or amino
end-groups. Results were obtained using the SIESTA implementation
of DFT \cite{Sol02}, which employs norm-conserving
pseudopotentials and linear combinations of atomic orbitals
\cite{SIES}. To treat exchange and correlation we used the
generalized gradient approximation (GGA) \cite{Per96}, but we also
made some tests with the local density approximation (LDA)
\cite{Per81}, which allowed us to check the influence of the
energy functional on structural and transport properties
\cite{LDAGGA}. Transport properties were calculated with the
SMEAGOL code \cite{Roc06}, which is interfaced to SIESTA and
self-consistently computes the charge density, the
electron-transmission coefficients and the I-V characteristics
\cite{SMEA}.

\begin{table}
\caption{Angles between rings calculated using LDA and
GGA.}\label{Tab01}
\begin{ruledtabular}
\begin{tabular}{ccccc}
\multicolumn{1}{c}{Molecule}&\multicolumn{2}{c}{Amines}&
\multicolumn{2}{c}{Thiols}\\
&LDA&GGA&LDA&GGA\\
\hline
2&0.0&0.0&0.0&0.0\\
3&15.6&14.9&14.7&15.7\\
4&29.4&31.5&30.4&30.9\\
5&44.1&52.5&43.2&46.8\\
6&50.4&57.6&50.8&49.3\\
7&54.0&61.2&62.5&62.8\\
8&87.6&89.5&89.4&88.1\\
\end{tabular}
\end{ruledtabular}
\end{table}

To find the most stable structure we started from a twisted
configuration, with an angle between 0 and 90 degrees, and then
allowed the molecule to freely relax. Since the presence of local
minima could have prevented the rings from rotating to the most
stable configuration, we used various initial angles to ensure
that we reached the absolute minimum. $\theta =0$ corresponds to a
meta-stable configuration for most of the molecules. For other
starting angles we found that the molecules always relaxed to the
most stable configuration. Table (\ref{Tab01}) shows the
ground-state angles calculated with LDA and GGA. Both functionals
give similar results, although in most cases the LDA
underestimates the angle. Both sets of results compare well with
previous GGA values, computed in Ref. \onlinecite{Ven06}. Table
(\ref{Tab01})also shows the dependence of $\theta$ on the
end-groups and reveals that the end group can change $\theta$ by
up to 15\%.

We used the relaxed conformation of the pristine molecule as a
starting point to compute the transport characteristics between
fcc (111) gold electrodes. We assumed that the protecting
hydrogens of the amino and thiol groups were lost when the
molecules were attached and the angle between the molecule and the
surface was 90 degrees. We carefully determined the distance
between the molecule and the gold leads by studying the energy as
a function of the distance of a simplified molecule composed of a
benzene attached to only one N or S on top of a (111) gold slab
made of 5 layers. With this distance and the molecule in its
ground-state conformation, we calculated the transmission
coefficient $T(E)$ for electrons of energy $E$ passing through the
device.

\begin{figure}
\includegraphics[width=\columnwidth]{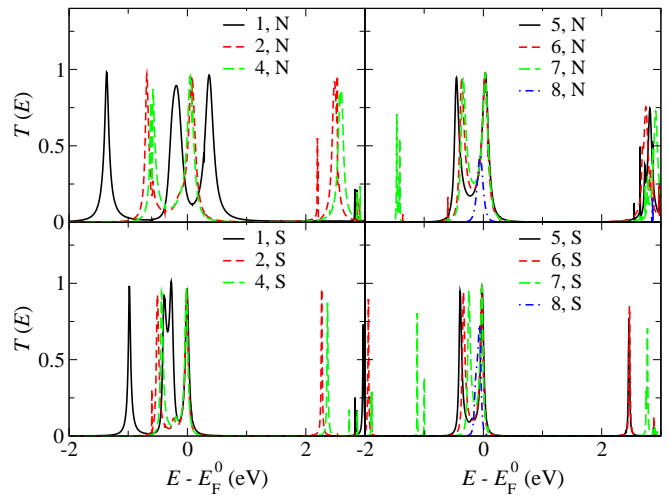}
\caption{\label{Fig2}Transmission curves for molecules 1-8 between
gold leads contacted by a surface gold adatom.}
\end{figure}

\begin{figure}
\includegraphics[width=\columnwidth]{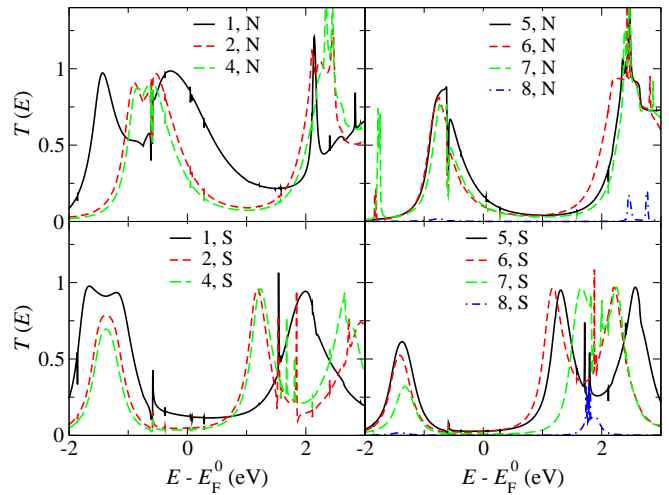}
\caption{\label{Fig3}Transmission curves for molecules 1-8 between
gold leads contacted by a three-atoms hollow site.}
\end{figure}

The transmission coefficients of all molecules between gold (111)
electrodes computed with GGA are shown in Figs. (\ref{Fig2}) and
(\ref{Fig3}) \cite{Molec3}. We found that nitrogen termination
hole-dopes the system, compared to the thiol termination, since
the former moves the HOMO closer to the computed Fermi energy
$E_\mathrm{F}^0=0$. We also studied two possible contact
geometries, namely a hollow position (Fig. (\ref{Fig3})), where
the contact atom sat on top of a 3-atoms hollow of the (111)
surface, and a top position (Fig. (\ref{Fig2})), where a gold
adatom was placed on top of the hollow position and in contact
with the molecule. By comparing both contact configurations we
found that the top position produces a much weaker coupling
\cite{Bra03}, manifested by the fact that the transmission
coefficients in Fig. (\ref{Fig2}) exhibit a series of sharp
Breit-Wigner resonances of height equal to unity. In the hollow
position, Fig. (\ref{Fig3}) shows that the Breit-Wigner
description breaks down, because the coupling is much stronger and
consequently, the resonances are merged and smeared out. The use
of LDA and/or DZP changes the height and shape of the LUMO peak,
but gives a similar value for the $T(E)$ at the Fermi energy and
almost the same HOMO peak.

Close inspection of the structure of $T(E)$ reveals information
about the molecular states. Fig. (\ref{Fig2}) shows the evolution
of the peaks in the top configuration as $\theta$ increases and
reveals that both HOMO peaks, which initially are clearly
separated, move together and eventually merge for angles close to
90 degrees. This behavior is reminiscent of bonding and
antibonding states with a coupling matrix element, which depends
on the overlap between the $\pi$ conjugated states of both rings.
From the same figure, it is also clear that the width of the
resonances does not decrease as the angle changes, which means
that the couplings to the leads (the $\Gamma$ matrices) are the
same for all molcules and the observed changes have an
intra-molecular origin.  In the hollow configuration, Fig.
(\ref{Fig3}) shows that $T(E)$ is characterized by the presence of
two broad peaks associated to HOMO and LUMO levels. As $\theta$
increases, the width of the HOMO decreases, which is again due to
the fact that the splitting between bonding and anti-bonding
states, within the HOMO peak, decreases with increasing $\theta$.

Having examined $T(E)$ for a range of energies, we now turn to the
low-voltage, low-temperature conductance
$G(\theta)=G_0T(E_\mathrm{F},\theta)$, where $G_0=2e^2/h$,
$E_\mathrm{F}$ is the Fermi energy and for ease of notation we
have now made the dependence of the transmission coefficient on
$\theta$ explicit. In figures (\ref{Fig2}) and (\ref{Fig3}), the
zero of energy is chosen to coincide with the computed Fermi
energy $E_\mathrm{F}^0$. In an experiment, $E_\mathrm{F}$ may
differ from $E_\mathrm{F}^0$ for a number of reasons, including
the presence of a dielectric environment, such as air or water
\cite{Che06,Lon06}. Furthermore, even in the absence of
environmental effects, differences can arise from the absence of
self-interaction corrections in the exchange-correlation
functional \cite{Toh06}. For these reasons, we shall treat
$E_\mathrm{F}$ as a free parameter and examine the $\theta$
dependence of $G$ for a range of $E_\mathrm{F}$.

\begin{figure}
\includegraphics[width=\columnwidth]{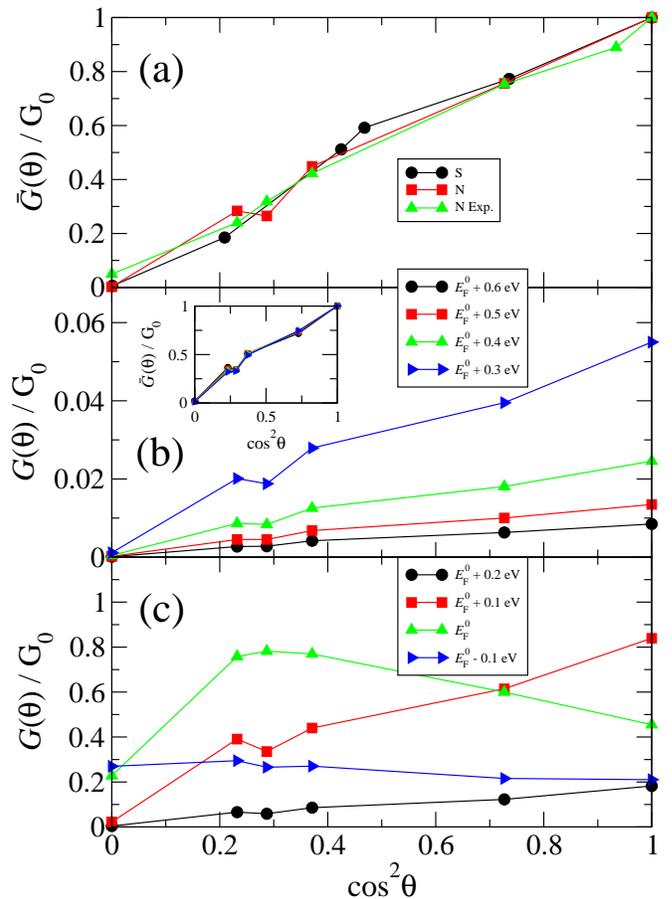}
\caption{\label{Fig4}(a) Zero-bias conductance in the hollow
configuration, for sulfur contact (circles), nitrogen contact
(squares) and values from Ref. \onlinecite{Ven06} (triangles). All
cases have been normalized to the $\theta=0$ value. (b) and (c)
computed low-bias conductance $G$ as a function of
$\mathrm{cos}^2\theta$ and the Fermi level for the the top
configuration and nitrogen contact. The inset shows the curves of
graph (b) normalized to the $\theta=0$ value.}
\end{figure}

Our first observation is that when the Fermi energy is inside the
HOMO-LUMO gap, the conductance is approximately a linear function
of $\mathrm{cos}^2\theta$, in agreement with the experiments of
Ref. \onlinecite{Ven06}. This result is illustrated in Fig.
(\ref{Fig4}) (a), which shows a comparison between experimental
results for the normalised conductance $\bar G=G(\theta)/G(0)$ and
theoretical results \cite{Norm}, obtained in the hollow
configuration, for both N and S couplings. (In the experimental
curve we also use our GGA-computed angular values.) As can be seen
the, agreement with the experimental curve is rather good. When
LDA is used, slight differences in the computed the angle can
shift the position of the points in the graph, but since the
conductance values are also different at those angles, the
dependence is still linear and the curve sits on top of those
shown in Fig. (\ref{Fig4}a).

Fig. (\ref{Fig4}) (b) shows plots of $T(E_\mathrm{F},\theta)$
obtained in the top configuration, for several values of
$E_\mathrm{F}$. Again we find a linear dependence on
$\mathrm{cos}^2\theta$, when the Fermi energy sits in the
HOMO-LUMO gap. If these curves are normalized to the $\theta=0$
value, as shown in the inset of Fig (\ref{Fig4}b), all of them sit
on top of each other, which indicates the robustness of the linear
dependence inside the HOMO-LUMO gap.

In contrast with the above behaviour, when $E_\mathrm{F}$
approaches either the HOMO or the LUMO orbitals, the $\theta$
dependence of the molecular resonances destroys the linear
behavior and $T(E_\mathrm{F},\theta)$ can even become
non-monotonic.  This is illustrated in Fig. (\ref{Fig4}) (c) for
the case of N bonding to a top site. Similar behavior is found in
the hollow configuration or when N is substituted by S as bonding
atom. This implies that by gating \cite{Kub03} the molecules, the
measured $\theta$ dependence could be tuned to be either linear or
non-monotonic, depending on the value of $E_\mathrm{F}$.

In summary, we have studied the transport properties of a series
of molecular wires as a function of their conformal, chemical and
contact configuration. We find that geometry is the dominant
factor when the Fermi level sits inside the HOMO-LUMO gap, since
different side groups do not affect the linear dependence.
However, when the Fermi level approaches either the HOMO or LUMO
resonances, chemistry comes into play, because the chemical
composition of the pristine molecules causes the positions of the
HOMO and LUMO resonances to differ from molecule to molecule. This
result shows that experiments such as those in Ref.
\onlinecite{Ven06} yield important spectroscopic information about
molecules in contact with electrodes, since the
experimentally-measured linear dependence on
$\mathrm{cos}^2\theta$, allows one to conclude that the Fermi
energy in these experiments lies in the HOMO-LUMO gap.

\begin{acknowledgments}
We thank Ian Sage and Martin Bryce for useful discussions and the
NWGrid for computing resources. We acknowledge financial support
from the European Commission and the British EPSRC, DTI, Royal
Society, and NWDA. CMF thanks QinetiQ for funding through
Agreement No. CU016-48671. VMGS thanks the EU network
MRTN-CT-2004-504574 for a Marie Curie grant.
\end{acknowledgments}


\begin{thebibliography}{99}
\bibitem{SIES} Since the geometrical configurations and lattice
constants depend strongly on the type of basis set we employed
double zetas and polarization orbitals (DZP) to perform the
structural relaxations. However, due to the fact that the
transmission coefficients and the value of the conductance at the
Fermi level converge very fast with the basis set we used a
single-zeta (SZ) to obtain the transport characteristics, but we
also made some tests with DZP. The Hamiltonian, overlaps and
electronic densities were evaluated in a real space grid defined
with a plane wave cutoff of 200 Ry. The coordinates were relaxed
until all forces were smaller than 0.02 eV/\AA\/.
\bibitem{LDAGGA} The optimized molecular geometries can be slightly
different depending on the exchange-correlation functional, with
the LDA/GGA tending to underestimate/overestimate distances and
angles. However, the general shape of the transmission
coefficients around the Fermi level and the zero-bias conductance
are very similar in both cases.
\bibitem{SMEA} The system was divided in three parts: left lead,
right lead and extended molecule (molecule plus part of the leads
modified by the presence of the molecule and the surfaces). The
unit cell of the leads was made of three slices of fcc gold grown
along (111) with 9 atoms per slice. In the extended-molecule
calculation we had to include one of the previous unit cells on
each side plus three and four additional slices on the left and
right sides of the molecule, respectively, to make the system
periodic and avoid surface problems. Each transport calculation
was composed then of around 150 atoms. We also used periodic
boundary conditions and 4 k-points along the perpendicular
directions to avoid spurious peaks and gaps in the transmission
coefficients.
\bibitem{Molec3} Molecule 3 overlaps strongly along the
perpendicular directions with molecules of adjanted unit cells due
to its large transversal size. This affects dramatically the
transmission curves and produces deviations from the expected
behaviors. For this reason we have decided not to include the
transport results of this molecule.
\bibitem{Norm} The dissagreement between theory and experiments
comes mainly from an incorrect determination of the contact
geometry \cite{Di00} and coupling distances \cite{Gar06} and can
also be amplified by the experimental enviroment
\cite{Che06,Lon06}. But since the highest resistance and therefore
the main transport characteristics are determined by what happens
in the central molecular region, the qualitative behavior and
general trends generated by intramolecular changes are correctly
described by theory.
\end{thebibliography}
\end{document}